\documentclass[reprint,IOP,superscriptaddress,amsmath,amssymb,article]{revtex4-2}
\usepackage{bm}        
\usepackage{amssymb}   
\usepackage{color}
\usepackage{units}
\usepackage{amsmath}
\usepackage{natbib}
\usepackage{esint}
\usepackage{chemformula}
\usepackage{ulem}
\usepackage[english]{babel}
\usepackage[colorlinks=true, citecolor=blue]{hyperref}
\usepackage{xcolor}
\usepackage{chemformula} 
\usepackage[T1]{fontenc} 
\usepackage{tikz}
\usepackage{parskip}
\usepackage{multirow}
\usepackage{graphicx}
\usepackage{mathtools} 
\usepackage{soul}
\usepackage{amsmath}

\DeclareUnicodeCharacter{2009}{\,}

\newcommand{\SPINX}{\affiliation{Spin-X Institute, School of Physics and Optoelectronics, State Key Laboratory of Luminescent Materials and Devices, Guangdong-Hong Kong-Macao Joint Laboratory of Optoelectronic and Magnetic Functional Materials, South China University of Technology, Guangzhou 511442, China}}

\newcommand{\PKU}{\affiliation{School of Physics, Peking University, Beijing 100871, China}}

\newcommand{\PKUM}{\affiliation{School of Materials Science and Engineering, Peking University, Beijing 100871, China}}
  
\newcommand{\SYSU}{\affiliation{School of Materials, Shenzhen Campus of Sun Yat-Sen University, Shenzhen 518107, China}}
\begin{document}

\title{Colossal
magnetoresistance in a quasi-two-dimensional cluster glass semiconductor}
\author{Suman Kalyan Pradhan}
\email{suman1kalyan@scut.edu.cn}
\SPINX

\author{Weiqi Liu}
\SPINX

\author{Jicheng Wang}
\SPINX

\author{Yongli Yu}
\SPINX

\author{Wenxing Chen}
\SPINX

\author{Jinbo Yang}
\PKU

\author{Yanglong Hou}
\SYSU
\PKUM

\author{Rui Wu}
\email{ruiwu001@scut.edu.cn}
\SPINX

\begin{abstract}
With a surge of interest in spintronics, the manipulation and detection of colossal magnetoresistance 
in quasi-two-dimensional layered magnetic materials have become a key focus, driven by their relatively scarce occurrence compared to giant magnetoresistance 
and tunneling magnetoresistance. This study presents an investigation into the desired colossal magnetoresistance, achieved by introducing magnetic frustration through Te doping in quasi-two-dimensional antiferromagnet Cr$_2$Se$_3$ matrix. The resulting  Cr$_{0.98}$SeTe$_{0.27}$ exhibits cluster glass-like behavior with a freezing temperature of 28 K. Magnetotransport studies reveal a significant negative magnetoresistance of up to 32\%. 
Additionally, 
angle-dependent transport measurements demonstrate a magnetic field-induced transition from positive to negative resistance anisotropy, suggesting a magnetic field-driven alteration in the electronic structure of this narrow band gap semiconductor, a characteristic feature of the colossal magnetoresistance effect. This behavior is further corroborated by density functional theory calculations. This systematic investigation provides a crucial understanding of the control of colossal magnetoresistance in quasi-two-dimensional materials via competing exchange interactions. 
 \end{abstract}
\maketitle
\section{Introduction}

The discovery of ferromagnetism (FM) in two-dimensional (2D) materials has significantly advanced our understanding of magnetism in reduced dimensions \cite{Gong2017DiscoveryOI}. This breakthrough has not only unveiled novel physical properties \cite{Yu2010RealspaceOO,PhysRevX.5.031013,Ali2016} but also introduced intriguing new physics unique to lower-dimensional systems. Consequently, extensive research in condensed matter and materials physics has focused on identifying pristine 2D van der Waals (vdW) layered magnetic materials \cite{Lee,Huang,Zhou,PhysRevB.108.115122}. 
Among the numerous material platforms engaged in searching, the recently emerged 2D magnetic semiconductors offer a wide array of remarkable functionalities, spanning electronic to magnetic properties, thanks to their nontrivial band topology \cite{Wang2018,Cai,Takahashi}.

Despite this progress, several key characteristics of these layered materials remain elusive. Among them, the glassy magnetic state and the giant negative magnetoresistance (NMR) are particularly intriguing. Magnetic glassiness often stems from inhomogeneity, frustration, or disorder among competing magnetic interactions, potentially giving rise to exotic magnetic states \cite{Glasbrenner2015,Khatua2023}.
On the other hand, negative magnetoresistance (NMR) is commonly observed in disordered or doped low-dimensional systems \cite{Bai2019} and serves as a hallmark of unique physical phenomena, including giant magnetoresistance (GMR) and tunneling magnetoresistance (TMR). 

In particular, some materials exhibit a distinct type of NMR, where the application of a magnetic field significantly reduces electrical resistance. This phenomenon, known as colossal magnetoresistance (CMR), makes these materials ideal for applications in magnetic sensors and spin field-effect transistors. Initially, CMR was attributed to an electronic phase transition, specifically the transition from antiferromagnetic (AFM) to ferromagnetic (FM) order, accompanied by a shift from an insulating or semiconducting state to a metallic one. In contrast, GMR and TMR originate from spin-dependent scattering, which varies between parallel and antiparallel magnetization alignments in different layers. La$_{1-x}$Ca$_x$MnO$_3$ is a prototypical material in which CMR was first reported. In this system, the double-exchange interaction between Mn$^{3+}$ and Mn$^{4+}$, combined with Jahn-Teller distortion, drives the insulator-to-metal transition, leading to CMR \cite{Salamon}. Similarly, Ti$_2$Mn$_2$O$_7$, despite lacking both double-exchange interactions and Jahn-Teller polarons, undergoes an insulator-to-metal transition and exhibits CMR \cite{Shimkawa}. Despite the structural and electronic differences between perovskite and pyrochlore manganites, a key characteristic common to CMR materials is the presence of an insulator-to-metal transition. More recently, CMR has also been observed in several non-oxide low-dimensional materials, such as EuMnSb$_2$ \cite{Sun2021}, EuTe$_2$ \cite{PhysRevB.104.214419}, and EuCd$_2$P$_2$ \cite{PhysRevB.108.L241115}, where an insulator-to-metal transition similarly occurs. Recent studies have also highlighted that an increasing number of material systems have been identified, each exhibiting distinct mechanisms underlying CMR \cite{PhysRevB.103.L161105,Seo2021,Wang2021,PhysRevB.101.205126,PhysRevB.104.L020408}. However, the number of quasi-2D layered materials exhibiting CMR remains very limited \cite{Zhang21,Zhu2022}. Therefore, understanding the fundamental mechanisms underlying CMR not only provides insight into the emergence of novel electronic phases but also paves the way for designing new 2D materials. In this context, Cr$_x$(Se$_y$/Te$_z$) compounds have attracted significant interest due to their diverse physical properties and exceptional tunability of magnetic and electronic states \cite{Zhang2019, Wu_2020, BoLi2021, Zhang2021, Zhong2023, Chen2023, 10.1063/5.0231254}. 

This communication presents the striking observation of CMR in two distinct layered Te-doped Cr$_2$Se$_3$ compounds, each exhibiting different origins for the phenomenon, synthesized using the chemical vapor transport (CVT) method. 
The formation of these compositions is confirmed by powder X-ray diffraction (P-XRD) refinement, which reveals that they share the same crystal structure. The composition Cr$_{0.98}$SeTe$_{0.27}$, 
exhibits cluster glass behavior due to competing magnetic exchange interactions below the freezing temperature of 28 K. This glassy behavior is validated through AC susceptibility measurements and a series of equilibrium dynamic analyses. Subsequently, the material exhibits a high colossal magnetoresistance of 32\%, while retaining its narrow bandgap semiconducting nature. Additionally, the CMR phenomenon is manifested through a field-induced change in the electronic structure, as reflected in the sign change of the angle-dependent magnetoresistance (ADMR) with the applied field. 
In contrast, the Te-excess composition is a ferromagnetic metal that exhibits reduced CMR, attributed to a semiconductor-to-metal transition at low temperatures.
\section{Results and Discussion}
\begin{figure*}
\centering
\includegraphics[width=2\columnwidth]{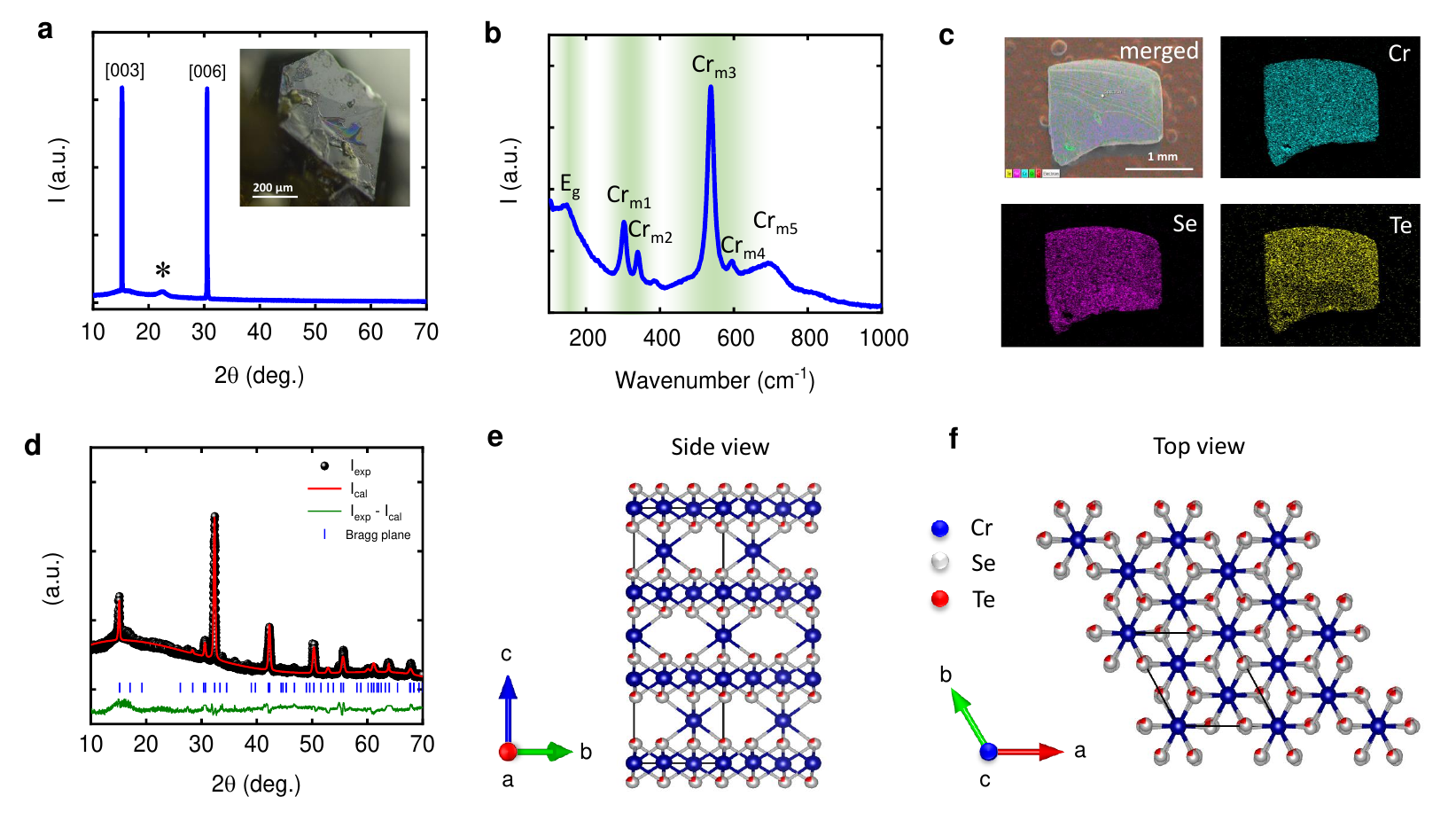}
\caption{\textbf{Preliminary characterization of as-grown crystal :} (a) Typical x-ray diffraction pattern obtained from the cleaved plane of a single-crystal at room temperature. The peak around 22.50$^{\circ}$ (marked as $\star$) corresponds to the background. An optical
micrograph image of the grown crystal is displayed in the upper inset. (b) Room-temperature in-plane Raman spectra. (c)
The area elemental color mapping of Cr, Se, and Te in green, violet, and yellow
color. The scale bar is 1 mm. (d) Experimental powder X-ray diffraction pattern (represented by solid black spheres) and simulated patterns (represented by a solid red line). The blue dashed lines represent the Bragg positions, while the green curve indicates the difference between the observed and simulated patterns. Rietveld refinement reveals that the XRD patterns correspond to a trigonal crystal structure with the $R$$\overline{3}$ space group (No. 148). (e,f) Display of the solved crystal structures from both the side and top views. The blue, gray, and red spheres represent Cr, Se, and Te atoms, respectively. The black box with cross-sectional rectangles indicates the crystallographic unit cell.}
\vspace{-0.45cm}
\label{cryst}
\end{figure*}
 
The study begins with crystal synthesis, followed by direct current (DC) and alternating current (AC) magnetic measurements, field- and angle-dependent magnetotransport measurements, and concludes with band structure calculations.

Bulk crystals were synthesized using the CVT method, resulting in mm-sized, shiny black crystals [inset of Fig.~\ref{cryst}(a)]. The details of the crystal growth process using CVT are provided in the Experimental Section of the Supporting Information. Fig.~\ref{cryst}(a) shows the typical x-ray diffraction (XRD) pattern of a bulk crystal. Featured intensity peaks of the [00${l}$] Bragg plane, indicate that the crystal grows along the $c$ axis. Moreover, the enlarged view of the [006] plane illustrates a single sharp peak having a full width at half maximum (FWHM) of $\Delta$$\theta$ = 0.09$^{\circ}$ 
(not shown here), indicating good crystallinity of the grown crystals. Room-temperature Raman scattering measurements reveal the vibrational modes at 147, 304, 342, 547, 596, 700 cm$^{-1}$ [Fig.~\ref{cryst}(b)], which correspond to $E_\text{g}$ (Te) \cite{Du2017}, $Cr_\text{m1}$, $Cr_\text{m2}$, $Cr_\text{m3}$, $Cr_\text{m4}$, and $Cr_\text{m5}$ \cite{Wu_2020} respectively. Energy dispersive X-ray (EDX) measurements confirm that the actual chemical composition is Cr$_{0.98}$SeTe$_{0.27}$ [Fig. S2]. The elemental color map is shown in Fig.~\ref{cryst}(c). The crystal structure was further probed by powder X-ray diffraction, performed on powdered samples obtained by grinding single crystals with well-characterized compositions. The refinement of the P-XRD data [Figs.~\ref{cryst}(d)] confirms that the grown crystal adopts a trigonal crystal symmetry in the $R$$\overline{3}$ space group (No. 148) with lattice parameters $a$ = $b$ = 6.25 $\mathring{A}$, $c$ = 17.28 $\mathring{A}$, $\alpha$ = $\beta$ = 90$^{\circ}$, and $\gamma$ = 120$^{\circ}$. The refined lattice parameters and atomic positions are provided in Tables S1-S2 in the Supporting Information.  
Figs.~\ref{cryst}(e,f) illustrate the corresponding schematic crystal structure from the side and top. The plane along $a$ is rectangular, while the plane exhibits a hexagonal arrangement (honeycomb) when viewed along the $c$ axis. 
We synthesized an additional composition using the same method, but with an excess of Te, while maintaining a Cr-to-Se ratio similar to that of the previous composition. Detailed characterization reveals that this composition exhibits similar crystalline symmetry [see Fig. S6(a)].\\


First, we investigated the magnetic properties of bulk  Cr$_{0.98}$SeTe$_{0.27}$ by measuring the temperature dependence of magnetization under a DC magnetic field. Measurements were carried out along the $c$ axis (out-of-plane, OOP) and the ${ab}$-plane (in-plane, IP) using the zero field-cooled (ZFC) and field-cooled (FC) protocols. Figure \ref{magnt}(a) illustrates the temperature ($T$) dependence of the magnetization $M(T)$ for bulk  Cr$_{0.98}$SeTe$_{0.27}$, measured with $H_\text{dc}$ = 0.1 kOe along the IP and OOP directions. The FC curve gives an initial impression of a ferromagnetic ground state, on account of both measuring directions. In contrast, the ZFC data, measured under similar $H_\text{dc}$ conditions, exhibit a cusp-like feature. 
 In addition, a substantial difference is observed between the ZFC and FC curve 
 just below the ordering temperature. 
 This pronounced splitting and cusp-like ZFC curve at temperatures below 30 K could hint at a robust glassy magnetic 
 state \cite{Pradhan2020}. 
 A much larger magnetization in the OOP direction than the IP direction indicates a strong magnetocrystalline anisotropy with an easy axis along the OOP direction. Additionally, the peak observed in the ZFC data, which is highly dependent on the applied magnetic field ($H_\text{dc}$), shifts to lower temperatures as $H_\text{dc}$ increases [see Fig.~\ref{magnt}(b)]. (The significance of this observation is discussed in a later section.) 
 
Next, we apply a \textit{modified Curie–Weiss} fit ($\chi$ = $\chi_o$ + ${C}$/{$T$-${\theta}$) to
the linear portion of the magnetic susceptibility in the temperature range 120 to 300 K, as shown in Fig.~\ref{magnt}(c). This analysis yields a Curie-Weiss temperature, $\theta$ = -14.66 K and -17.50 K, and the effective magnetic moment, $\mu_\text{eff}$ = 3.12 $\mu_\text{B}$ and 3.101 $\mu_\text{B}$ in bulk Cr$_{0.98}$SeTe$_{0.27}$ along IP and OOP respectively. The negative $\theta$ observed here indicates significant antiferromagnetic coupling in this FM-like compound. The effective magnetic moments closely match the Cr${^{+3}}$ atomic moment, indicating that the magnetic moment is primarily contributed by Cr. Along with this, the frustration parameter ($F$), which quantifies the degree of frustration in the magnetic system, can be calculated as: |$F$| = $\theta_\text{CW}$/$T_\text{M}$, where $T_\text{M}$ is the peak temperature observed under zero-field cooling. In this case, $F$ is approximately 0.52, indicating moderate frustration in the composition \cite{Ramirez}. 

		%

Figure \ref{magnt}(d) shows the ZFC $M(H)$ measurements conducted at $T$ = 2 K, with a magnetic field applied up to $\pm$ 50 kOe along the IP and OOP directions. The data reveal a wide hysteresis loop with coercive fields $H_\text{C}$ of 3.85 kOe for IP and 4 kOe for OOP. Additionally, the magnetization $M$ displays a non-saturating tendency, increasing almost linearly with increasing $H_\text{dc}$. 
These characteristics indicate the presence of a mixed magnetic ground state at low temperatures. Several studies have reported on the exchange bias (EB) effect, in mixed-phase magnetic materials with competing magnetic interactions \cite{Pradhan2020} so far. However, the identical zero-field-cooled and field-cooled hysteresis loops measured at 2 K, shown in the inset of Fig.~\ref{magnt}(d), indicate that the exchange bias effect is not present here. Similar to the 
$M(T)$ behavior, the $M(H)$ data also indicate an easy axis of magnetic anisotropy along the out-of-plane direction. 
\begin{figure}
\centering
\includegraphics[width=1\columnwidth]{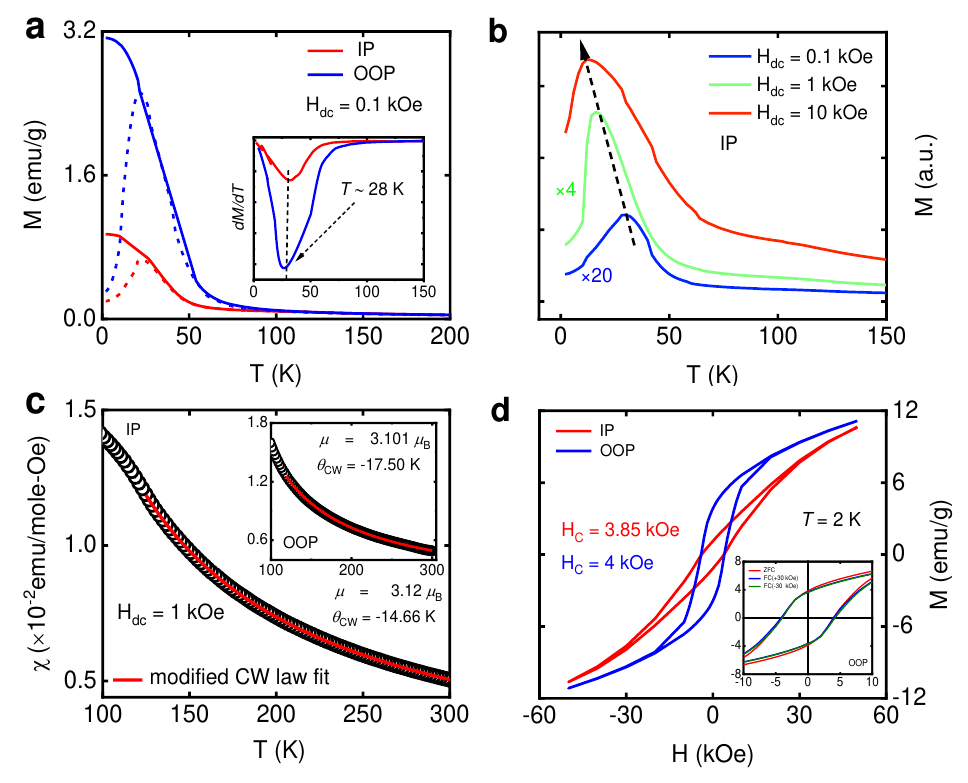}
\caption{\textbf{Preliminary DC magnetic characterization :} (a) Temperature-dependent magnetization $M$($T$) measured under ZFC and FC modes with a magnetic field $H_\text{dc}$ = 0.1 kOe. 
Scattered and continuous lines, respectively represent ZFC and FC curves. The corresponding inset shows the plot ${dM}$/${dT}$ vs. $T$. (b) Displays the ZFC $M$($T$) curves at different $H_\text{dc}$. (c) Modified Curie-Weiss fit (in solid red line) for
$H_{\text{dc}}$ = 1 kOe. 
The inset shows the parameters obtained. (d) The isothermal magnetization curve measured at $T$ = 2 K, 
and the corresponding coercivities ($H_\text{C}$) are mentioned in the lower inset. The inset indicates the absence of the exchange bias effect. Measurements were carried out along IP and OOP directions respectively. }
\vspace{-0.65cm}
\label{magnt}
\end{figure}
\\\\
The observation of field-dependent ZFC peak maxima in Fig. \ref{magnt}(b), 
as discussed earlier, suggests several possible spin configurations. These may include a chiral helimagnetic (CHM) state \cite{Han2017}, 
or a glassy magnetic phase \cite{Pradhan2020}. To validate the potential spin configuration as a chiral helimagnetic (CHM) state, a few characteristics need to be confirmed: (i) the presence of an in-plane magnetic easy axis and (ii) the observation of multistep metamagnetic transitions in the $M(H)$ curves at different temperatures \cite{Kurumaji2017}. The ratio
$\ \frac{M_\text{OOP}}{M_\text{IP}}$ $\approx$ 4 (for $H_\text{dc}$ = 1 kOe and $T$ = 2 K) obtained from Fig. \ref{magnt}(a) indicates a significant magnetic anisotropy in this compound, with the magnetic easy axis aligned along the out-of-plane direction ($c$-axis). This is further supported by the $M(H)$  data at $T$ = 2 K, as shown in Fig. \ref{magnt}(d). Additionally, Fig.~\ref{magnt}(d) shows a single-step magnetization process, with $M$ increasing linearly up to the maximum applied field, ruling out the presence of multi-step magnetization transitions. Therefore, the bulk crystal does not exhibit characteristics typical of CHM states. 

To examine the contribution of spin clusters, we measured the AC susceptibility ($\chi$ = $\chi$$\prime$ + $\chi$$\prime\prime$, 
where 
$\chi$$\prime$ represents the reversible magnetization process and $\chi$$\prime\prime$ 
accounts for losses due to irreversible magnetization processes). 
These measurements were conducted with an oscillating AC field ($H_\text{ac}$) of 4 Oe at various frequencies ($f$) in the temperature range from 80 K to 2 K along the IP and OOP directions. As shown in Fig.~\ref{acx111}(a,b), a prominent and well-defined peak appears in both the real part ($\chi$$\prime$) and the imaginary part ($\chi$$\prime\prime$) of the susceptibility data, closely corresponding to or very near the ordering temperature observed in the DC magnetization curves. The peak in $\chi$$\prime$(${f,T}$) shifts to higher temperatures and its amplitude decreases with increasing frequency [Fig.~\ref{acx111}(a,b)]. This behavior confirms the presence of a glassy magnetic phase, characterized by slow spin dynamics. This is in contrast to ferromagnets and conventional magnetic systems, where such shifts in $\chi$$\prime$(${f,T}$) typically occur in the MHz frequency range \cite{Irkhin1999}. Similarly to $\chi$$\prime$(${f,T}$), the imaginary part of the susceptibility 
$\chi$$\prime\prime$(${f,T}$) also shows a distinct peak near the glassy transition, as indicated in the insets of Fig.~\ref{acx111}(a,b). The position of this peak shifts to higher temperatures with increasing frequency and the amplitude of $\chi$$\prime\prime$ decreases to zero after reaching the peak. These characteristics are indicative of slow magnetic relaxation \cite{Cortie2020} in the compound.

To categorize glassy magnetic systems based on the response of magnetic spins and the relative shift in freezing temperature with frequency, a quantitative approach known as the Mydosh parameter ($K$) is employed \cite{Mulder1981}.

\begin{center}
	\begin{equation}
		\begin{split}
			K=\frac{\bigtriangleup T_{pf}}{T_p\bigtriangleup(log(f))}  \\
		\end{split}
	\end{equation}
	\label{eq:Mydosh}
\end{center}
where $T_\text{p}$ is the temperature at which the peak of the AC susceptibility occurs, and $f$ is the measured frequency. Here, 
$\bigtriangleup T_\text{pf}$ = $T_\text{Pf1}$-$T_\text{Pf2}$ and $\bigtriangleup log(f) = log(f1)-log(f2)$. The Mydosh parameter, $K$ obtained here is 0.036 for IP and 0.037 for OOP measurements. This value best corresponds to an intermediate situation, indicative of a cluster glass (CG) type system (\textit{K} $\leq$ 0.08) \cite{V.K.2020}. Generally, CG systems are understood as ensembles of interacting ferromagnetic clusters arranged randomly in the antiferromagnetic matrix.

The frequency dependence of $T_\text{p}$ follows the conventional power-law divergence associated with critical slowing down (CSD),
	\begin{center}
		\begin{equation}
			\begin{split}
				\tau=\tau_0 (\frac{T_p-T_{f}}{T_f})^{-{z\nu^\prime}}  \\
			\end{split}
		\label{eq:powerlaw}
		\end{equation}
	\end{center}

where $\tau$ = relaxation time corresponding to the measured frequency($\tau=1/\nu$), $\tau_0$ = characteristic relaxation time of the individual spin cluster, $T_\text{f}$ = freezing temperature for $\nu\rightarrow$ 0 Hz, and ${z\nu^\prime}$ = dynamic critical exponent [$\nu^\prime$ = critical exponent of correlation length, $\xi= (T_\text{p} /T_\text{f} -1)^{-\nu^\prime}$ and  $\tau\thicksim\xi^z$] \cite{Mydosh2015}.	Equation~\ref{eq:powerlaw} can be rewritten as, $ln(\tau)$ = $ln(\tau_0)$ - ${z\nu^\prime}ln(t)$,
where $t$ = $\frac{T_\text{p}-T_\text{f}}{T_\text{f}}$. The plot of ln($\tau$) versus \textit{ln}($t$) is shown in Fig.~\ref{acx111}(c,d). From the slope and intercept we can estimate the value of ${z\nu^\prime}$ and $\tau_0$ respectively. The obtained values of the characteristic relaxation $\tau_0$, and ${z\nu^\prime}$ are $2.06\times 10^{-9}$ s, $7.58\times 10^{-10}$ s, 12.418 $\pm$ 2.126 and 15.271 $\pm$ 1.152  for IP and OOP respectively. These results of $\tau_0$ are comparable with a disordered CG system \cite{V.K.2020}, further suggesting slow spin dynamics in Cr$_{0.98}$SeTe$_{0.27}$ due to strongly interacting clusters. This is further corroborated by the \textit{Vogel-Fulcher}(VF)  relaxation model, as detailed in the Supporting Information. On the other hand, the Te-excess composition exhibits ferromagnetic (FM) behavior with an ordering temperature of approximately 66 K and a markedly lower degree of magnetic frustration [see Fig. S6(b-d)].\\

\begin{figure}
\centering
\includegraphics[width=1\columnwidth]{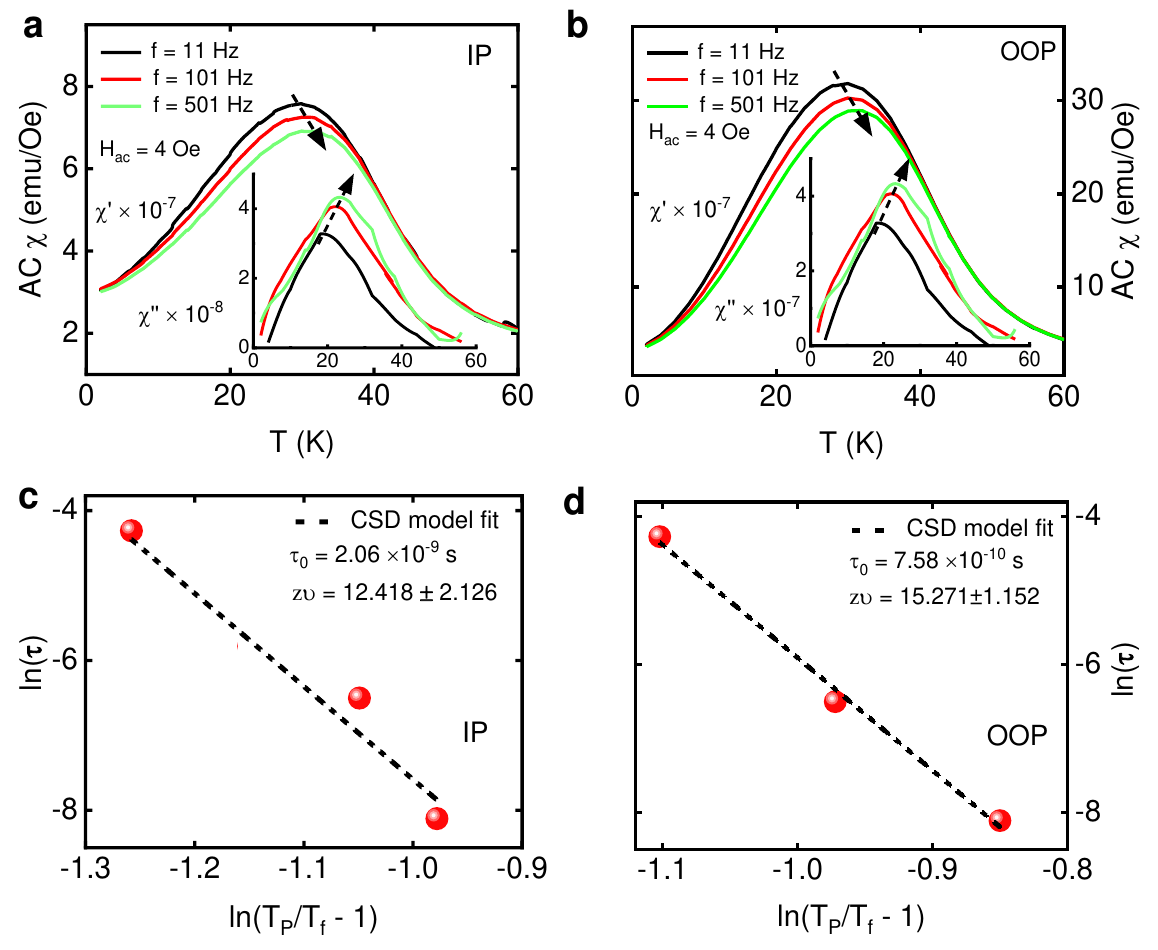}
\caption{\textbf{Temperature dependence of AC-susceptibility data measured at various excitation frequency 
:} Temperature response of real part of AC susceptibility  $\chi$$\prime$($T$) measured at $H_\text{ac}$ = 4 Oe along (a) IP and (b) OOP at different measuring frequencies. Corresponding insets depict the temperature dependents of $\chi$$\prime\prime$($T$). The arrow indicates the variation in the maximum value of AC susceptibility ($T_\text{P}$) with increasing frequency. (c,d) Represent the CSD model fit. Fitted parameters are also mentioned in the figures.}
\vspace{-0.45cm}
\label{acx111}
\end{figure}

Having established Cr$_{0.98}$SeTe$_{0.27}$ as an intriguing magnetic material, we now proceed to characterize its electronic transport 
response to the external field. To study the transport properties of this bulk crystal, the in-plane resistivity ($\rho\textsubscript{xx}$) was measured using a linear four-probe configuration, as shown in Fig. S1(b). Fig.~\ref{Rxx}(a) exhibits the temperature-dependent electrical resistivity 
at $H$ = 0. With increasing temperature, $\rho\textsubscript{xx}$ decreases monotonically throughout the $T$ scan, implying a semiconductor-like nature.  The resistivity at 5 K, corresponding to the residual resistivity, is $\sim$ 0.17 m$\Omega$cm. The residual resistivity ratio, defined as, $\rho\textsubscript{xx}$(300 K)/$\rho\textsubscript{xx}$(5 K), is 0.294, which is comparable to that of the well-known spin-gapless semiconductor Mn$_2$CoAl \cite{Ouardi2013}. Circumstantially, the behavior of $\rho\textsubscript{xx}$ at $H$ = 60 kOe is strikingly similar to that at $H$ = 0 [Fig.~\ref{Rxx}(a)], but with a
negative response to the external magnetic field, indicating the emergence of a negative MR effect. Inset of Fig.~\ref{Rxx}(a) displays a significant change in $\rho\textsubscript{xx}$, defined by $\Delta$$\rho\textsubscript{xx}$(\%) = [$\rho\textsubscript{xx}$($H$ = 60 kOe) - $\rho\textsubscript{xx}$($H$ = 0)]/$\rho\textsubscript{xx}$($H$ = 0) $\times$ 100, 
which occurs primarily at low temperatures 
(< 30 K). In addition, $\rho\textsubscript{xx}$ shows a pronounced change in slope near the magnetic ordering at 30 K [see Fig. S4(a)], which tends to smooth out upon the application of a magnetic field, suggesting suppression of spin fluctuations. 
However,  the magnitude and variation of resistivity 
with temperatures are relatively small compared to conventional semiconductors. This type of nontrivial semiconducting nature has been previously reported in some topological materials \cite{Jun,LiQ}. 
Furthermore, the Arrhenius fit 
in the temperature interval 85–300 K yields a low activation energy of
6.84 meV (see Fig.S3(a)), which implies a semiconductor with a narrow band gap.\\ 

These observations motivate us to study the magnetotransport property of these grown crystals. Therefore, we studied the isothermal variation of $\rho\textsubscript{xx}$($H$) in two configurations, 
$H$ 
$\parallel$ ${ab}$
and $H$ $\perp$ 
${ab}$  
from 5 to 60 K, up to an external field strength of 60 kOe. 
To proceed, we define the MRs as, 
MR(\%) = [$\rho$($H$) - $\rho$(0)/$\rho$(0)] $\times$ 100, where $\rho$($H$) and $\rho$(0) are resistivities under an applied magnetic field and without a magnetic field, respectively. The MR has been symmetrized with respect to positive and negative magnetic fields to eliminate the Hall contribution.

Significant NMR is observed throughout all measurements below 100 K [see Fig.~\ref{Rxx}(b)], 
expected for a disordered system. Finding NMR is quite common for 
3$d$ transition metal-based intermetallic alloy showing FM-like
order, due to the suppression of the spin-disorder scattering with increasing magnetic field \cite{LiPi2002,PhysRevB.66.024433}. As the temperature increases, the suppression of spin scattering may diminish, leading to a relatively small NMR. 
The main outcome is the observation of a fairly large NMR reaching $\sim$ 32\% under 60 kOe at 5 K [more preciously $\sim$ 8\% at 10 kOe, which is known as the low-field MR (LFMR)]. Thereafter, the magnitude decreases with increasing $T$. This can be comparable to the MR values of some giant NMR materials like bulk YCuAs$_2$ \cite{Kang}, and Fe/Cr multilayers \cite{Gijs}. The variation of 
pronounced NMR is well displayed in Fig.~\ref{Rxx}(b) and S5(a); with decreasing $T$ along two measuring configurations. 
Qualitatively, the nature of MR along $H$ $\parallel$ ${ab}$ 
and $H$ $\perp$ ${ab}$ 
are found to be almost identical since it does not depend on the orientation of the magnetic
field relative to the crystal surface, (thereby ruling out any possible chirality). In addition, a butterfly-shaped MR 
could be observed at $T$ = 5 K.  This noticeable MR hysteresis is found to be consistent with the $M(H)$ curve as shown in Figure S3(b), and further signifies a weak FM fluctuation at low $T$. Normally, a butterfly-shaped MR is believed to be an important characteristic of an FM conductor.
Upon further warming up, the butterfly-shaped smears out. Eventually, the MR loop appears in a sharp cusp shape and finally becomes broad, 
indicating a magnetic phase transition occurs below 35 K. Together with Fig.~\ref{Rxx}(b) and S5(a) also depict the MR response taken at 5 K and 60 K for 0$^\circ$ and 90$^\circ$ indicate no possible spin-flop transition. 
In $H$ $\perp$ 
${ab}$ configuration 
at 60 K, MR is found to obey a dependence of nearly $H$ (predicted in $s$-$d$ scattering model), while MR shows a  $H^{2/3}$ variation below $T_\text{C}$ [Fig. S5(a)]. This nearly $H^{2/3}$ dependence of MR around $T_\text{C}$, is found to be similar to that predicted in a few theoretical works \cite{doi:10.1143/JPSJ.34.51,PhysRevB.66.024433}. In contrast, MR follows a nearly quadratic and linear field dependency above and below the ordering temperature along $H$ $\parallel$ 
${ab}$ [Fig. S5(b)]. 

\begin{figure}
\centering
\includegraphics[width=1\columnwidth]{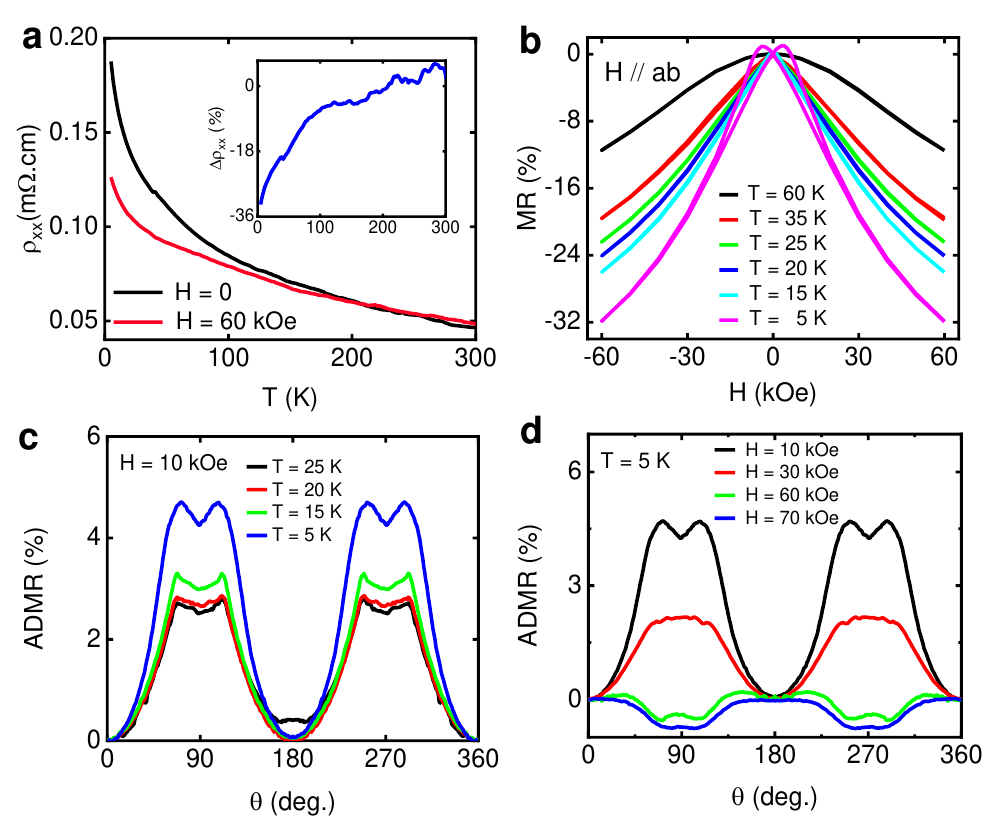}
\centering
\caption{\textbf{Electric transport properties :} (a) Temperature-dependent in-plane resistivity $\rho_{xx}$($T$) at $H$ = 0, and 60 kOe exhibiting
a semiconductor-like behavior. The upper right insets show the temperature dependence of the difference magnetoresistance under 0 and 60 kOe
magnetic fields. 
(b) In-plane field-dependent MR measured as a function of $H$ in a temperature range of 5 K to 60 K in $H$ $\parallel$ ${ab}$ configurations. 
(c) Angular dependence of
MR (ADMR) under an in-plane magnetic field of 10 kOe at different temperatures. 
(d) ADMR measured for various applied fields at 5 K.} 

\vspace{-0.45cm}
\label{Rxx}
\end{figure}
Significant NMR in a material is often termed colossal magnetoresistance, but several conditions must be fulfilled for this phenomenon to manifest. As shown in Fig.~\ref{Rxx}(a), $\rho\textsubscript{xx}$ shows only a minimal change in magnitude at low temperatures in response to the application of a magnetic field. In contrast, well-known CMR materials exhibit drastic decreases in $\rho\textsubscript{xx}$, often by 5–7 orders of magnitude, at low temperatures \cite{Seo2021}. The coinciding magnetic and resistivity transitions observed here are typical in CMR materials \cite{Jiang2006}. 
This implies that the observed NMR cannot merely be attributed to spin-scattering effects, but rather, it stems from electronic phase transitions that are intimately linked to the field-induced magnetic phase transitions within this material. Therefore, these observations position the present composition as an intriguing CMR material. In contrast, the Te-excess composition exhibits metallic behavior, featuring a semiconductor-to-metal transition $\sim$ 25 K, and demonstrates reduced colossal magnetoresistance of 12\% [see Fig. S6(e,f)].

To explore the anisotropy in magnetoresistance and further validate the CMR, we performed angle-dependent magnetoresistance (ADMR) measurements at different temperatures and magnetic fields, with the angle $\theta$ varied from 0$^\circ$ to 360$^\circ$. 
As for these measurements, the magnetic field was kept perpendicular to the current direction. The measurement setup geometry is clearly illustrated in Fig. S1(c). As shown in Fig.~\ref{Rxx}(c), we observe the prominent ADMR [$\rho$($\theta^\circ$) - $\rho$($0^\circ$)/$\rho$(0$^\circ$) $\times$ 100] with an almost perfect four-fold oscillation for the lowest measuring $T$ at $H$ = 10 kOe ($\sim$ 5\% at 5 K). The small ADMR \cite{Ahadi} in magnetic materials primarily arises from anisotropic spin-dependent scattering due to spin-orbit coupling (SOC) \cite{Shick}. 
From Fig.~\ref{Rxx}(c), it can be easily found that the minimum in ADMR appears whenever $\theta$ = 0$^{\circ}$, 
180$^{\circ}$, 
360$^{\circ}$, with two saddle points at 90$^{\circ}$ and 270$^{\circ}$. More importantly, the curve is symmetric concerning 180$^{\circ}$. 
The polar plot of Fig.~\ref{Rxx}(c) [not shown here] displays a butterfly-like pattern in the ADMR at 5 K. As the temperature increases towards magnetic ordering, ADMR is gradually suppressed but remains detectable compared to the ADMR at 5 K, implying a close relation between the ADMR and the magnetic order of Cr moments. 
These observations [Fig.~\ref{Rxx}(c)] indicate a hidden two-fold symmetry also exists along with four-fold symmetry, indicating highly anisotropic Fermi surface characteristics.  
At 5 K, the twofold component dominates over the fourfold component.  However, as the temperature increases, the fourfold component dies faster than the twofold component, implying that the cause of the fourfold component is distinct from that of the twofold component. Detailed analysis can be found in the Supporting Information.

The field dependence of ADMR at 5 K exhibits dramatic changes in shape and magnitude, as illustrated in Fig.~\ref{Rxx}(d). Notably, the ADMR is positive for $H$ = 10, and 30 kOe, but negative for $H$ = 60 kOe, and 70 kOe, highlighting its sensitivity to both the field angle and the magnitude of the applied magnetic field. The polar plot of ADMR presented in Fig. S4(d) clearly demonstrates that the shape evolves from a butterfly-like form to a dumbbell-like one, with the positions of the maxima and minima shifting as the applied field changes. Magnetic field–induced changes in the ADMR symmetry are unconventional. For example, in Sr$_2$IrO$_4$, where magnetoelastic effects have been proposed as the underlying cause \cite{Wang2014}. The Dirac semimetal ZrTe$_5$ exhibits similar characteristics; however, the underlying cause is different. A topological quantum phase transition involving the field-induced annihilation of Weyl points is the likely explanation \cite{Zheng2017}. However, our experimental observations rule out these possibilities, suggesting instead a significant change in the electronic band structure with the magnetic field [positive ADMR $\rightarrow$ negative ADMR at 60 kOe, as shown in Fig.~\ref{Rxx}(d)]. The sign change in ADMR with $H$ suggests the presence of spin fluctuation scattering in this system. Typically, such a phenomenon occurs within a specific doping range, highlighting the importance of the underlying electronic structure \cite{Ahadi2018}.\\



To investigate the mechanism of the CMR effect in this present compound, it is useful to revisit various mechanisms discussed above \cite{Sun2021,PhysRevB.105.165122,PhysRevB.101.205126,PhysRevB.104.L020408,PhysRevB.108.205140}. However, these mechanisms are not applicable in this case. Instead, the origin can be understood in terms of competition between different magnetic phases \cite{Wildman2012,PhysRevLett.123.047201}, as indicated by the presence of a glassy magnetic state in the system. For instance, chromium and selenium-based Cr$_2$Se$_3$ is a distinct 2D layer-structured antiferromagnet, which feature a noncollinear spin configuration with weak ferromagnetic moments \cite{
Wu_2020},
 while Cr$_x$Te$_z$ are primarily 2D layered ferromagnetic metal \cite{Zhang2021,Zhong2023,Chen2023,
10.1063/5.0231254}. Specifically, in this case, the bulk crystal system exhibits a frustration parameter of 0.52, indicating moderate geometric frustration, which gives rise to this complex magnetic ground state [see Fig.~\ref{DOS}(a)]. 
Therefore, when a magnetic field is applied, it promotes ferromagnetic ordering [Fig.~\ref{DOS}(b)], which on one hand reduces spin-dependent scattering of charge carriers, and on the other hand, induces an electronic structure change tending towards metallic behavior. This, in turn, leads to a further decrease in resistance as the magnetic field increases, resulting in the colossal magnetoresistance observed in  Cr$_{0.98}$SeTe$_{0.27}$. Whereas the CMR in the Te-excess composition is attributed to a semiconductor-to-metal transition at low temperatures \cite{Sun2021}.

\begin{figure}
\centering
\includegraphics[width=1\columnwidth]{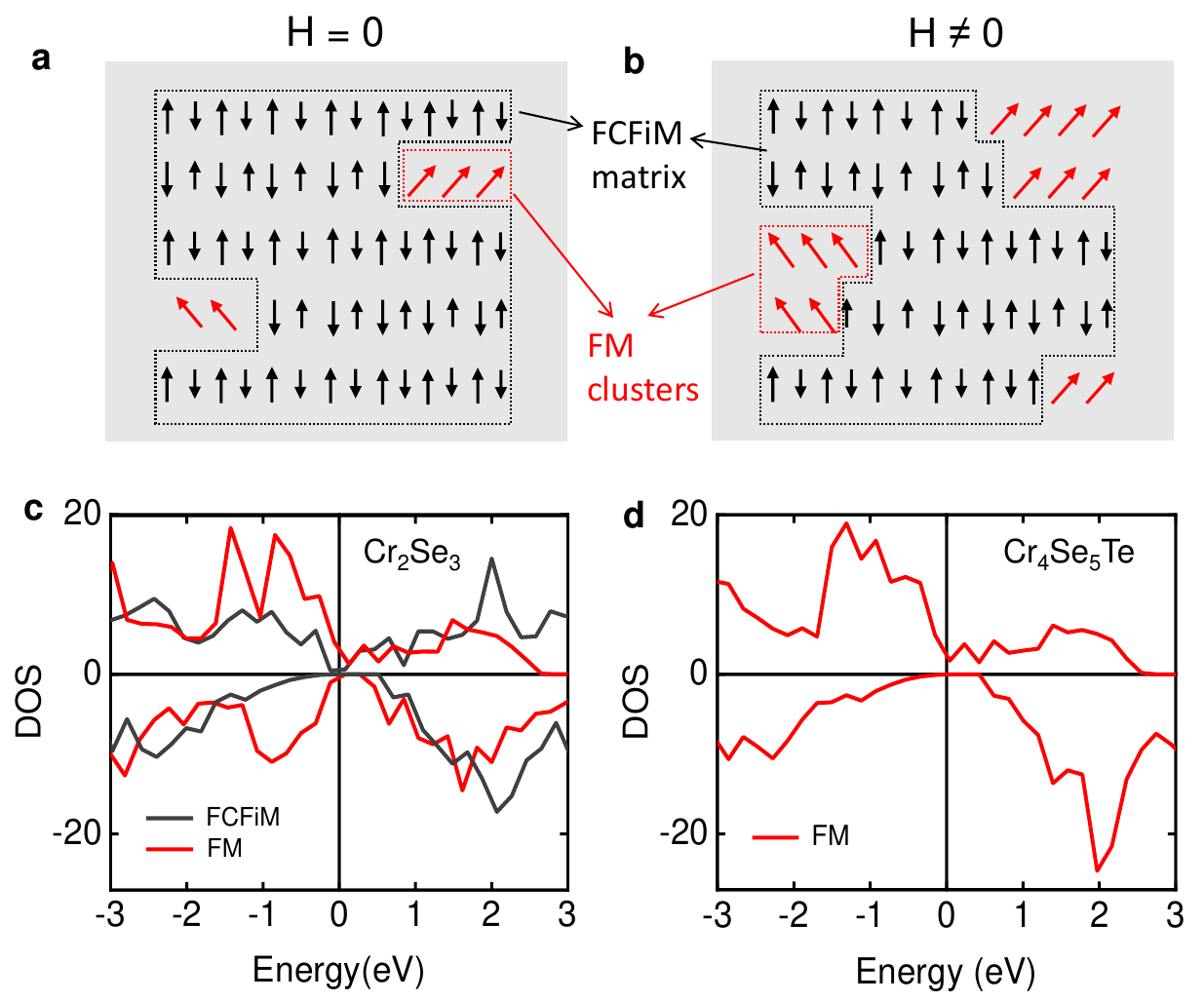}
\caption{\textbf{Theoretical interpretation of experimental results :} A simplified schematic illustration of the inhomogeneous magnetic phase (a) for $H$ = 0, and (b) for $H \neq$ 0. Here, the arrows denote the direction of the spins. The regions marked with black and red arrows correspond to the FCFiM matrix and FM clusters, respectively. Band structure of the (c) Cr$_2$Se$_3$, and (d) Cr$_4$Se$_5$Te compositions.}
\vspace{-0.45cm}
\label{DOS}
\end{figure}
 We performed density functional theory (DFT) calculations to further support the infusion of strong ferromagnetic interactions in Cr$_2$Se$_3$ through Te doping. For these calculations, we considered the trigonal crystal structures of both the pristine and doped materials. To determine the magnetic ground state of Cr$_2$Se$_3$, we consider four possible magnetic configurations: ferromagnetic (FM), ferrimagnetic (FiM-I and FiM-II), and fully compensated ferrimagnetic (FCFiM). As summarized in Table 1, our calculations reveal that the magnetic ground state of Cr$_2$Se$_3$ is the FCFiM state, in agreement with previous theoretical study \cite{Gebredingle}. Next, to investigate the effect of Te substitution, we replace a Se atom in the primitive cell with a Te atom. The resulting Cr$_4$Se$_5$Te crystal adopts a ferromagnetic ground state, indicating a magnetic phase transition upon Se-to-Te substitution, which aligns with our experimental observations. The density of states (DOS) for both systems are shown in Fig.~\ref{DOS}(c,d). Both Cr$_2$Se$_3$ and Cr$_4$Se$_5$Te exhibit half-metallic behavior, with Cr$_4$Se$_5$Te displaying a larger band gap. Additionally, Cr$_2$Se$_3$ in the FM state shows a larger band gap compared to its magnetic ground state.

\begin{table}[h!]
	\caption{$\Delta$E represents the energy difference between various magnetic states and the magnetic ground state. The arrows indicate the spin orientations of different Cr sites.} 
	\centering 
	\begin{tabular}{|p{2.3cm} p{1.7cm} p{2.3cm} p{1.7cm}|}
		\hline
		Cr$_2$Se$_3$  & $\Delta$E (MeV) & Cr$_4$Se$_5$Te & $\Delta$E (MeV) \\
		\hline
		FM (↑↑↑) & 9.48  & FM (↑↑↑) & 0  \\
		FiM-I (↑↓↑) & 30.61  & FiM-I (↑↓↑) & 38.72 \\
            FiM-II (↑↑↓) & 12.09  & FiM-II (↑↑↓)& 13.77  \\
            FCFiM (↓↑↑) & 0  & FCFiM (↓↑↑) &  10.84\\
				\hline
	\end{tabular}
    \label{SP}
\end{table}

\section{Outlook and Conclusion}
In summary, we have presented a comprehensive characterization of the magnetic and electronic properties of Cr$_{0.98}$SeTe$_{0.27}$ across its entire low-temperature phase diagram, complemented by first-principles calculations. The crystals were grown using the CVT method, and P-XRD refinement was employed to confirm the crystal structure. The crystal exhibits a cluster glass magnetic state below the freezing temperature of 28 K, which is due to competing ferromagnetic and antiferromagnetic interactions. A substantial colossal magnetoresistance of up to 32\% at 5 K is a key finding in this narrow-band gap semiconductor, independent of the direction of the magnetic field to the crystal surface. 
CMR is also evidenced by the magnetic field–induced sign change in the ADMR at 5 K. This CMR originates from the alteration in the electronic structure caused by the applied magnetic field. Additionally, the observed four-fold symmetry in the ADMR suggests a complex Fermi surface. 
Therefore, Cr$_{0.98}$SeTe$_{0.27}$ 
emerges as an intriguing 2D magnetic semiconductor with diverse and fascinating transport properties, offering a novel material platform for angle-sensitive spintronic applications that warrants further exploration.  On the other hand, the Te-excess composition is found to be a ferromagnetic metal with a semiconductor-to-metal transition at low temperatures, exhibiting reduced colossal magnetoresistance. Together, these findings suggest that increasing the Te doping concentration in Cr$_2$Se$_3$ alleviates geometric frustration, leading to a reduction in the CMR effect with a distinct underlying origin.


See the \textbf{Supporting Information} section for detailed experimental information, EDX analysis, P-XRD refinement parameters, and additional magnetic and transport results.

\section{Acknowledgement}
This work is supported by the National Key R\&D Program of China (grant no. 2022YFA1203902), the National Natural Science Foundation of China (NSFC) (grant nos. 12374108, 12104052, and 12241401), and the Guangdong Provincial Quantum Science Strategic Initiative (Grant No. GDZX2401002), the Fundamental Research Funds for the Central Universities, the State Key Lab of Luminescent Materials and Devices, South China University of Technology, and GBRCE for Functional Molecular Engineering. 

\textbf{Conflict of Interest:}
The authors declare that they have no conflict of interest.

\textbf{Data availability:} The data supporting the findings of this study are available from the corresponding author upon
reasonable request.

\appendix
\renewcommand{\thefigure}{A\arabic{figure}}
\renewcommand{\thesection}{A\arabic{section}}
\setcounter{figure}{0}
\setcounter{section}{0}

\end{document}